\documentclass[preprintnumbers, floatfix, twocolumn,
preprintnumbers, PRD, superscriptaddress,nofootinbib]{revtex4-1}
\usepackage{mathtools,mathptmx,array,slashed}
\usepackage{latexsym}
\usepackage{color}
\usepackage{float}
\renewcommand{\&}{\textup{\symbol{`\&}}}
\usepackage{hyperref}
\usepackage{booktabs}
\usepackage{graphicx}
\usepackage{subfigure}
\usepackage{epsf}
\usepackage{bm}

\newcommand{\be}{\begin{equation}}
\newcommand{\ee}{\end{equation}}

\newcommand{\mincir}{\raise
-3.truept\hbox{\rlap{\hbox{$\sim$}}\raise4.truept\hbox{$<$}\ }}
\newcommand{\magcir}{\raise
-3.truept\hbox{\rlap{\hbox{$\sim$}}\raise4.truept\hbox{$>$}\ }}

\def\be{\begin{equation}}
\def\ee{\end{equation}}
\def\bea{\begin{eqnarray}}
\def\eea{\end{eqnarray}}
\def\ba{\begin{aligned}}
\def\ea{\end{aligned}}

\providecommand{\U}[1]{\protect\rule{.1in}{.1in}}
\usepackage{tikz,xcolor,hyperref}

\begin{document}
\title{Universal Thermodynamic Topological Classes of Black Holes in a Perfect Fluid Dark Matter Background}

\author{Muhammad Rizwan}
\email{mrizwan@numl.edu.pk}
\affiliation{Department of Mathematics, Faculty of Engineering and Computing, National University of Modern Languages, H-9, Islamabad 44000, Pakistan}
\author{Mubasher Jamil}
\email{mjamil@sns.nust.edu.pk (corresponding author)}
\affiliation{
Department of Mathematics, School of Natural Sciences, National University of Sciences
 and Technology (NUST), H-12, Islamabad 44000, Pakistan
}
\author{M. Z. A. Moughal}
\email{zubair.moughal@ceme.nust.edu.pk}
\affiliation{
Department of Basic Sciences and Humanities, College of Electrical and Mechanical Engineering, National University of Sciences and Technology (NUST), H-12, Islamabad 44000, Pakistan
}

\begin{abstract}
In this paper, we study the universal thermodynamic topological classes of a family of black holes in a perfect fluid dark matter (PFDM) background. Recent research on black hole thermodynamics suggests that all black holes can be cast into four universal thermodynamic classes, denoted by $W^{1-}$, $W^{0+}$, $W^{0-}$, and $W^{1+}$. Our study reveals that the Schwarzschild black hole in PFDM belongs to the $W^{1-}$ class, and the independence of black hole size is thermodynamically unstable in both the low- and high-temperature limits. The Reissner–Nordström, Kerr, and Kerr–Newman black holes in the PFDM background belong to the same universal thermodynamic class, $W^{0+}$, which represents small-stable black holes and large-unstable black holes at low-temperature limits, whereas no black hole state exists at high temperatures. The AdS black holes behave differently when compared to their counterparts in PFDM. The Schwarzschild–AdS black hole belongs to the $W^{0-}$ class, indicating that no black hole state exists at low temperatures, but small-unstable and large-stable black hole states exist at high temperatures. Furthermore, the Kerr–AdS black hole belongs to the $W^{1+}$ class, characterized by small-stable black holes at low temperatures, large-stable black holes at high temperatures, and unstable intermediate-sized black holes at both low and high temperatures. Thus, the presence of PFDM does not affect the stability of the black hole. These findings uncover the universal topological classifications underlying black hole thermodynamics, offering profound insights into the fundamental principles of quantum gravity.
\end{abstract}

\maketitle

\section{Introduction}

The prediction of black holes by the theory of General Relativity has been widely tested experimentally, more notably via the Event Horizon Telescope (EHT) to observe shadows of supermassive black holes \cite{eht1,eht2}, as well as signature of gravitational wave generated via merger of black holes via LIGO/VIRGO observatories \cite{gw}. The astronomical observations suggest that black holes are surrounded by magnetic fields and plasma and actively accreting dust and plasma. One of the earliest proposals for the formation of black holes is through the collapse of a supermassive star (or a spherically symmetric gravitational collapse of pressureless dust) during the final stages of the star's evolution \cite{os,penrose}. However, black holes can also be formed by the merger of highly dense stars such as neutron stars \cite{ns}. Using observed candidates of black holes, the theories of modified gravity can be tested by confronting various black hole solutions in modified gravity theories with the astronomical observations. From theoretical perspective, novel features of black holes have been already studied. The profound connection between black holes and thermodynamics can be traced back to the groundbreaking work of Hawking \cite{hawkingbh}. Another important discovery is the interesting connection between the Einstein field equations, the fluid dynamics and thermodynamics \cite{pady}. In literature, charged and spinning black holes have been tested for stability under accretion of charged or spinning particles to validate the weak cosmic censorship conjecture as well \cite{sanjar,sanjar1}.

A novel approach for investigating the thermodynamic aspects of black holes has been proposed in \cite{Wei:2022dzw}, where the authors performed a comprehensive examination of black holes by studying the topological classes of black hole thermodynamics using the generalized off-shell free energy setup. As a result, the manifold of black hole solutions is categorized into three distinct topological classes, their differentiation arising from diverse topological numbers. Here, our focus is the examination of the thermodynamic topological categories pertaining to both stationary and axi-symmetric black holes immersed in PFDM. In literature, several physical aspects of black holes in PFDM background have been investigated including black hole shadows, quasinormal modes, and deflection angles \cite{pfdm0,pfdm1,pfdm2,pfdm3,pfdm4,pfdm5,rizwanatal}. The topological approach to study the black hole thermodynamics was used in recent works \cite{rtop1, rtop2PFDM, top1,top2,top3,top4,top5,top6} and, in particular, extended to rotating black holes in \cite{Wu:2022whe}. 
In what follows, we shall sketch a few concepts and provide an overview of the topological methodology introduced in \cite{Wei:2022dzw}. To proceed, the notion of the generalized off-shell Helmholtz free energy is given by 
\begin{equation}
\mathcal{F} = M -\frac{S}{\tau},
\end{equation}
where  $M$  and $S$ represent the mass and entropy, respectively, and $\tau$ is the inverse temperature parameter related to the cavity \cite{Wei:2022dzw}. The generalized Helmholtz free energy exhibits only on-shell properties and reduces to the standard Helmholtz free energy $F = M - TS$ of the black hole when $\tau = \beta = 1/T$ \cite{rtop3unipaper,rtop4unitopo}. 
Furthermore, as an integral component of this framework, a vector field $\phi$ which is gradient of the function can be introduced as \cite{Wei:2022dzw}
\begin{equation}\label{2}
    \phi = (\phi^{r_h},\phi^\Theta) =\Big(\frac{\mathcal{\partial \tilde{F}}}{\partial r_{h}}\, , ~\frac{\mathcal{\partial \tilde{F}}}{\partial \Theta}\Big) \, ,
\end{equation}
where the function $\tilde{\mathcal{F}}$ is related to the off-shell free energy $\mathcal{F}$, as 
\begin{equation}
    \tilde{\mathcal{F}}=\mathcal{F}+\frac{1}{\sin\Theta}.
\end{equation}
The two key parameters, $r_h$ and $\Theta$, are confined to the following domains $0 < r_h < +\infty$ and $0\le \Theta\le \pi$, respectively. It should be noted that the component $\phi^\Theta$ diverges at $\Theta = 0,\pi$, where it points an outgoing pointing vector in the respective regions. The point where $\phi^{r_h}$ vanishes identifies black hole states as zero points (or defects) of the vector field. Duan's $\phi$-mapping topological current theory suggests that a topological charge, known as the winding number $w$, can be assigned to each zero point or black hole state~\cite{duan1, duan2}. A positive winding number represents a positive heat capacity, which corresponds to a stable black hole, whereas a negative winding number corresponds to a negative heat capacity, which is responsible for an unstable black hole. The sum of the winding numbers, known as the topological number $W$, can be determined as follows:  
\begin{equation}
   W = \sum_{i=1}^{N}w_{i}\, 
\end{equation}  
where $w_i$ represents the winding number corresponding to the ith zero point and characterizes the thermodynamical topological class of the black hole.
Recent studies suggest that all black holes can be universally classified into four thermodynamical classes: $W^{1-}, ~W^{0+}, ~W^{0-}$, and $W^{1+}$, based on the stable and unstable black hole states indicated by the winding numbers of the vector field $\phi$~\cite{rtop3unipaper,rtop4unitopo}. This topological classification may provide insights into thermodynamic properties, contributing to a better understanding of black hole physics.

This paper is organized as follows: In sections II we do an examination of the topological classes associated Schwarzschild black holes in the presence of PFDM. We extend the analysis in section III to encompass the Reissner–Nordström black hole existing within the PFDM framework. In section IV, the topological classes of Kerr, and Kerr–Newman black are discussed. In section V and VI, we discuss how can the Schwarszchild-AdS and Kerr-AdS black holes be classified thermodynamically. Our paper culminates with a presentation of conclusions, outlined in section VII. We shall adopt units such that $c=1=G.$

\section{Schwarzschild Black Hole in a Dark Matter Background}

Astronomical observations and various phenomenological models of the galactic centers compellingly suggest that black holes are enveloped by dark matter \cite{jamil,jamil1}. The presence of dark matter ultimately has discernible and measurable effects on nearby objects, such as stars and the accretion disk.
In this section, we discuss the universal thermodynamic class of a (static) black hole in the PFDM background. The line element of the Schwarzschild black hole immersed in the PFDM background can be expressed as follows \cite{SBHDM, pfdm2}
\begin{eqnarray}\label{SBHPFDM}
ds^{2} &=&-f(r)dt^2+\frac{dr^2}{f(r)}+r^2(d\theta^2+\sin^2\theta d\phi^2),
\end{eqnarray}%
with
\begin{eqnarray}\label{fofBH}
f(r)=1-\frac{2m}{r}+\frac{\alpha}{r} \ln\left( \frac{r}{|\alpha |}\right),
\end{eqnarray}%
where $m$ and $\alpha$ represent the mass of the black hole, and the PFDM parameters, respectively. Notably, for every choice of the parameter $\alpha$, similar to a Schwarzschild black hole, the line element \eqref{SBHPFDM} describes a black hole with a single horizon, known as the event horizon, denoted by $r_h$. Thus, the presence of the PFDM background does not alter the number of horizons. However, the size of the event horizon increases with an increase in the dark matter parameter $\alpha$ \cite{rizwanatal}. The mass and thermodynamic quantities for the Schwarzschild black hole in the PFDM background are \cite{MJP}
\begin{equation}
    M= m,\quad S=\pi r_h^2, \quad  T=\frac{1}{4\pi r_h}\left(1+\frac{\alpha}{r_h}\right).
\end{equation}

Note that, in the absence of the PFDM, $\alpha=0$, the line element and the thermodynamical quantities reduce to that of the Schwarzschild black hole. For the Schwarzschild black hole in PFDM background the asymptotic behaviour of the inverse temperature parameter $\beta$ admits the following limits:
\begin{eqnarray}\label{boundary}
    \beta(r_m)=0, \quad \text{and} \quad \beta(\infty)=\infty,
\end{eqnarray}
where $r_m$ represents the minimal size of the black hole which is in this case is zero but for other black hole spacetimes it can be nonzero. Further, for universal thermodynamic analysis, the defect curve $\beta({r_h})$ is to be analytic in the range $(r_m, \infty)$. These limits provide valuable insights into the thermodynamical classification and stability of the black hole states.  

\begin{figure}
\centering
\includegraphics[width=7.5cm,height=7.5cm]{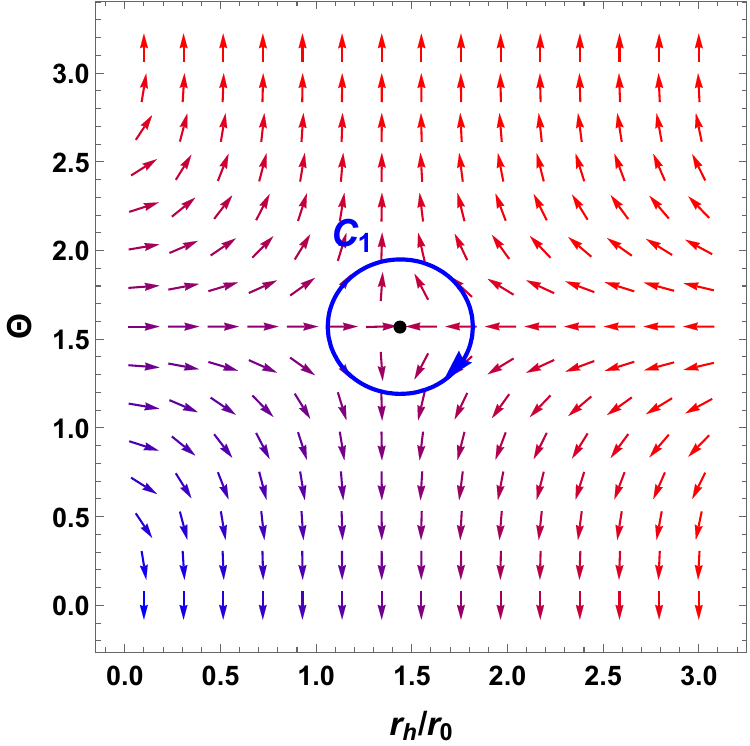}
\caption{The $n$-vector field $\phi$ is plotted on a portion of the $r_h$-$\Theta$ plane plotted for the Schwarzschild black hole in PFDM background with $\alpha=1/2$, and $\tau=4\pi$. The zero point of $\phi$ is marked with a black dot and is located at $(r_h,\Theta)=(1.44,\pi/2)$. The contour $C_1$ is closed loop enclosing zero point.}
\label{VFKiselevBH}
\end{figure}
\begin{figure}
\centering
\includegraphics[width=7.8cm,height=5.2cm]{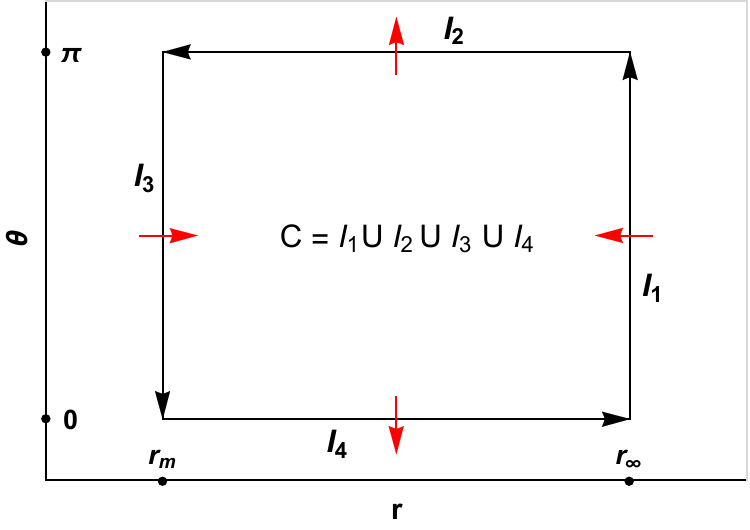}
\caption{To study the asymptotic behavior of the $n$-vector field $\phi$ at the boundary, we consider a contour $C = I_1 \cup I_2 \cup I_3 \cup I_4$ as shown here. The red arrows indicate the direction of the vector field $\phi$ for the Schwarzschild black hole in a PFDM background at the boundary, which encompasses all possible parameter regions. A similar boundary can be considered for all other black hole solutions to analyze the asymptotic behavior of the vector field.}
\label{boundSBHPF}
\end{figure}
\begin{figure}
\centering
\includegraphics[width=7.5cm,height=7.5cm]{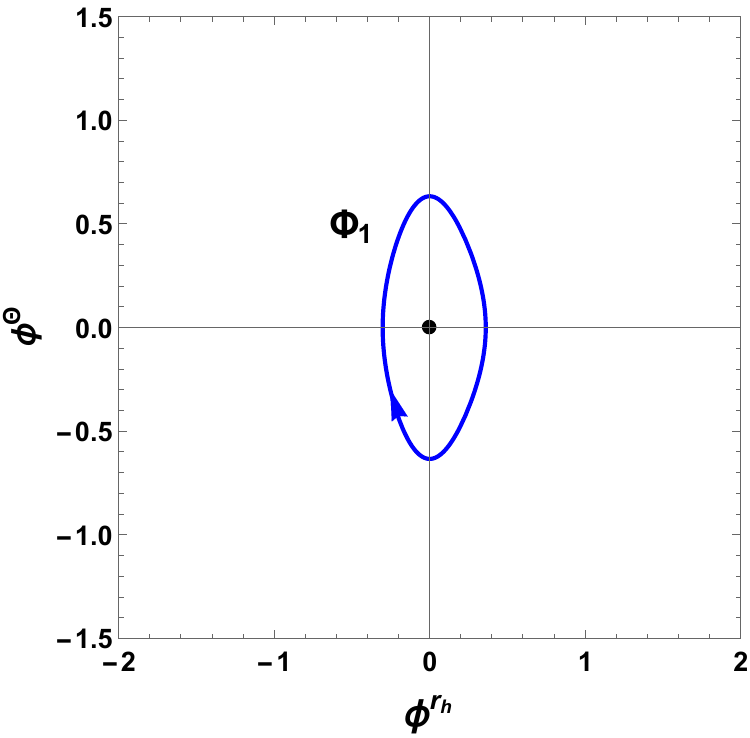}
\caption{Contour $\Phi_1$ represents the change in the components of the vector field $\phi$ as the contour $C_1$ in Fig. \ref{VFKiselevBH} is traversed for a Schwarzschild black hole in PFDM. The black dot indicates the zero point of the vector field, and the arrows show the sense of rotation that the vector $\phi$ undergoes while traversing the contour $C_1$. The changes in the components of the field along this closed loop cause $\Phi_1$ to rotate in a clockwise direction, corresponding to a winding number of $-1$.  }
\label{F2KiselevBHWN}
\end{figure}
The universal thermodynamical topological classification requires the new form of the off-shell free energy $\tilde{{\mathcal{F}}}$ of the black hole spacetime and the $n$-vector field $\phi$. The generalized off-shell free energy for the Schwarzschild black hole in PFDM background takes the form
\begin{eqnarray}\label{F}
\mathcal{F}&=&\frac{1}{2}\left[r_h+\alpha \ln\left( \frac{r_h}{|\alpha |}\right)\right]-\frac{\pi r_h^2}{\tau}.
\end{eqnarray}
The off-shell free energy allows us to define a new function incorporating an additional parameter, $\Theta$,
\begin{eqnarray}
\tilde{\mathcal{F}} = \frac{1}{2}\left[r_h+\alpha \ln\left( \frac{r_h}{|\alpha|}\right)\right] - \frac{\pi r_h^2}{\tau} + \frac{1}{\sin\Theta}.
\end{eqnarray}
The $n$-vector field $\phi$ is defined as the gradient of $\tilde{\mathcal{F}}$ {[see Eq. \eqref{2}]}, and its components for the Schwarzschild black hole in the PFDM background are given as:
\begin{eqnarray}
\phi^{r_h} &=& \frac{1}{2}\left(1+\frac{\alpha}{r_h}\right) - \frac{2\pi r_h}{\tau}, \\
\phi^\Theta &=& -\cot{\Theta} \csc{\Theta}.
\end{eqnarray}

Now, we study the behavior of the vector field $\phi$ (plotted in Fig. \ref{VFKiselevBH}) at the boundaries corresponding to Eq. \eqref{boundary}. The overall boundary is described by a contour $C = I_1 \cup I_2 \cup I_3 \cup I_4$, where the segments $I_i$ and the contour $C$ are shown in Fig. \ref{boundSBHPF}, encompassing all regions of the parameters. The setup of $\phi$ indicates that it is orthogonal to $I_1$ and $I_3$ \cite{rtop1}, therefore the key asymptotic behavior occurs along $I_2$ and $I_4$.
As $r_h \to r_m$, the vector field plot of $\phi$ suggests that the direction of $\phi$ is rightward with an inclination depending on $\Theta$, whereas when $r_h \to \infty$, it is leftward directed (see Fig. \ref{VFKiselevBH} and \ref{boundSBHPF}) suggesting it thermodynamical class.   

The Schwarzschild black hole in a PFDM background (similar to a Schwarzschild black hole) exists in only one state for a given $\tau$, characterized by a negative heat capacity and a topological number $W = -1$ \cite{rtop2PFDM}. This topological number can also be verified by studying the variations in the components of the vector field $\phi$ in the $\left(\phi^{r_h},\phi^\Theta \right)$ plane. The zero points of the vector field $\phi$, represented by black dots in the figures, are located at the origin, and tracking the changes of $C_1$ generates a closed loop $\Phi_1$ in the corresponding vector space (see Fig. \ref{F2KiselevBHWN}). Contour with a negative winding number are mapped to clockwise loop, indicating that the topological number is $W = -1$. 

The universal thermodynamical behavior of the Schwarzschild black hole in PFDM background can be analyzed as follows. For the limit $\beta \to \infty$, that is, in the low-temperature limit, the system features large black holes that are thermodynamically unstable due to their negative topological number. For the limit $\beta \to 0$, that is, for high temperatures, we have small black holes that are also unstable. Thus, we can conclude that, based on the universal thermodynamical class proposed in Ref. \cite{rtop3unipaper}, the Schwarzschild black hole in PFDM belongs to the class $W^{1-}$. 

\section{Reissner–Nordström black hole in dark matter background}

Now we investigate the impact of the electric charge in the topological numbers of thermodynamics. To find the spacetime geometry, one can proceed using the action for the gravity theory minimally coupled with gauge field in PFDM, which is read as \cite{Das:2020yxw}
\begin{eqnarray}
\mathcal{S}=\int dx^4\sqrt{-g}\left( \frac{1}{16\pi G} R+\frac{1}{4} F^{\mu \nu}F_{\mu \nu} + \mathcal{L}_{DM}\right).
\end{eqnarray}
In the last equation $g$ = det($g_{ab}$) is the determinant of the metric tensor, $R$ is the Ricci scalar, further $G$ is Newton's gravitational constant, $F_{\mu \nu}= \nabla_\mu A_\nu -\nabla_\nu A_\mu$ is electromagnetic field tensor and $\mathcal{L}_{DM}$ is the Lagrangian density for PFDM.  Upon using the variation for the action principle, one can obtain the Einstein field equations \cite{Das:2020yxw}
\begin{eqnarray}
R_{\mu \nu}-\frac{1}{2}g_{\mu \nu}R = 8\pi G (T_{\mu \nu}^M +T_{\mu \nu}^{DM}), 
\end{eqnarray}
along with
\begin{eqnarray}
F^{\mu \nu}_{;\nu}=0, \nonumber \\
F^{\mu \nu ;\alpha} +F^{\nu \alpha ;\mu}+F^{\alpha \mu ; \nu} =0.
\end{eqnarray} 
Above $T_{\mu \nu}^M$ and $T_{\mu \nu}^{DM}$ represent the energy-momentum tensor for ordinary matter and PFDM respectively,
 \begin{eqnarray}
T^\mu_\nu =g^{\mu \sigma} T_{\sigma \nu}, \nonumber \\
T^t_t = -\rho, \;\;\; T^r_r = T^{\theta}_{\theta} = T^{\phi}_{\phi} = P.
\end{eqnarray} 

It was shown that the line element of the Reissner–Nordström black hole  in the dark matter background is
given as \cite{Das:2020yxw}
\begin{eqnarray}\label{LEKNPFDM}
ds^2 &=&-f(r)dt^2+\frac{1}{f(r)}dr^2+r^2\left(d\theta^2+\sin^2\theta d\phi^2\right),
\end{eqnarray}
where
\begin{eqnarray}\notag
 f(r)=1-\frac{2m}{r}+\frac{Q^2}{r^2}+\frac{\alpha}{r}\ln\left(\frac{r}{|\alpha|}\right).
\end{eqnarray}
Here $Q$ is the electric charge of the black hole. The root analysis of the horizon equation $f(r)=0$ shows that for any choice of $\alpha$  if the charge parameter $Q$ are in the limit $Q<k_c$\footnote{The critical values is given as 
\begin{eqnarray}
        k_c=\frac{\alpha}{2}  \sqrt{W\left(2 e^{-1+\frac{2 M}{\alpha}}\right) \left[2+W\left(2 e^{-1+\frac{2 M}{\alpha}}\right)\right]}.
\end{eqnarray}
}, the line element \eqref{LEKNPFDM} represents a black hole with two horizons and a naked singularity otherwise \cite{rizwanatal}. For all choices of parameter $\alpha$, if the charge parameter satisfy the condition $Q=k_c$ the line element \eqref{LEKNPFDM} represent the extremal black hole with horizon given as 
\begin{eqnarray}\label{extremal_radius}
    r_e=\frac{\alpha}{2} W\left(2 e^{-1+\frac{2 M}{\alpha}}\right),
\end{eqnarray}
where $W(x)$ is the Lambert $W$-function. The thermodynamic quantities for this spacetime can be written as 
\begin{eqnarray}
    M&=&m, \quad S={\pi r_h^2}, \quad T=\frac{1}{4\pi r^3_h}(r^2_h+\alpha r_h-Q^2).
\end{eqnarray}
The asymptotic behaviour of the inverse temperature parameter $\beta$ for the Reissner–Nordström black hole in PFDM background has the following limits
\begin{eqnarray}
    \beta(r_m)=\infty \quad \text{and} \quad \beta({\infty})=\infty,
\end{eqnarray}
where $r_m=r_e$ with $r_e$ being the radius of extremal black hole given by Eq. (\ref{extremal_radius}).

\begin{figure}
\centering
\includegraphics[width=7.5cm,height=7.5cm]{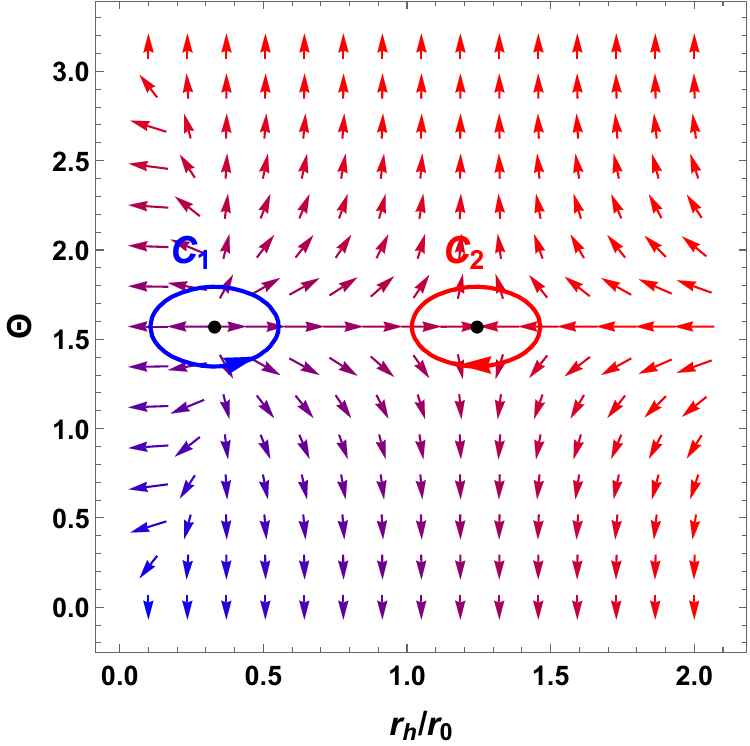}
\caption{The $n$-vector field on a portion of the $r_h$-$\Theta$ plane plotted for the Reissner–Nordström black hole in PFDM background with $\alpha=1/2$, $Q=1/2$ and $\tau=4\pi$. The zero points of $\phi$ are marked with black dots and are located at $(0.33,\pi/2)$ and $(1.24,\pi/2)$.}
\label{VFchKiselevBH}
\end{figure}
\begin{figure}
\centering
\includegraphics[width=7.5cm,height=7.5cm]{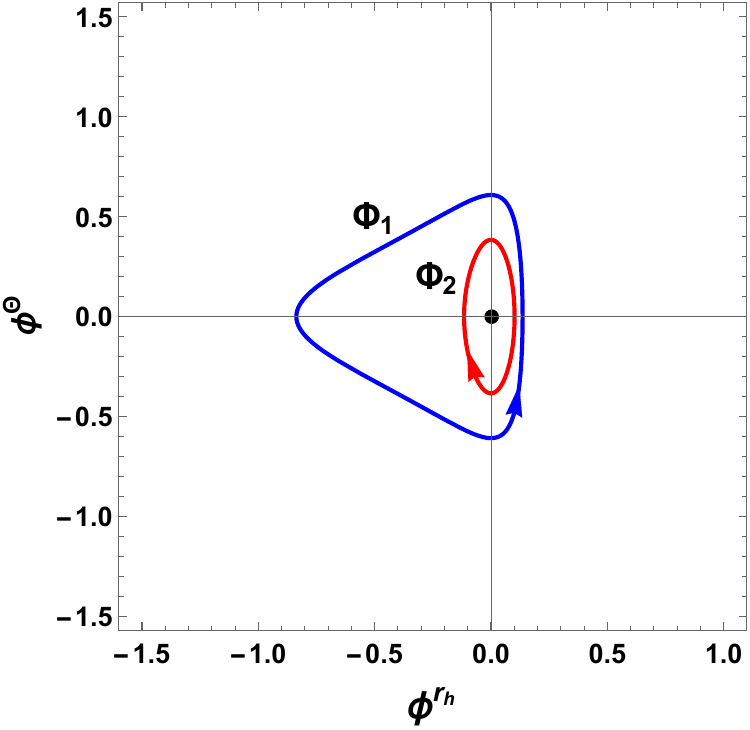}
\caption{Contours $\Phi_1$ and $\Phi_2$ represent the changes in the components of the vector field $\phi$ as the contours $C_1$ and $C_2$ in Fig. \ref{VFchKiselevBH} are traversed for a Reissner–Nordström black hole in PFDM. The black dots indicate the zero points of the vector field, and the arrows show the direction of rotation of the vector $\phi$ while traversing the contours $C_1$ and $C_2$. The changes in the components of the field along the closed loop cause $\Phi_1$ to rotate in a counterclockwise direction, corresponding to a winding number of $1$, whereas along loop $\Phi_2$, it rotates in a clockwise direction, corresponding to a winding number of $-1$.}
\label{F2ChKiselevBH}
\end{figure}

The generalized off-shell free energy of the Reissner–Nordström black hole in PFDM background  can be obtained as
\begin{eqnarray}\label{FRN}
\mathcal{F}&=&\frac{1}{2}\left[r_h+\frac{Q^2}{r_h}+\alpha \ln\left( \frac{r_h}{|\alpha |}\right)\right]-\frac{\pi r_h^2}{\tau},
\end{eqnarray}
and the corresponding components of the $n$ vector field $\phi$ are obtained as
\begin{eqnarray}\label{phiRN}
\phi^{r_h} &=& \frac{1}{2}\left(1+\frac{\alpha}{r_h}-\frac{Q^2}{r^2_h}\right) - \frac{2\pi r_h}{\tau}, \\
\phi^\Theta &=& -\cot{\Theta} \csc{\Theta}.
\end{eqnarray}
The $n$-vector field $\phi$ for the Reissner–Nordström black hole in the PFDM background is plotted in Fig. \ref{VFchKiselevBH}, and it asymptotic behavior at the boundaries $I_i$, as discussed for the Schwarzschild black hole in the PFDM background (in Fig. \ref{boundSBHPF}), is summarized in Table 1. It is observed that as $r_h \to r_m$ or $r_h \to \infty$, the vector field plot of $\phi$ suggests a lefward direction suggesting it universal thermodynamical class.   

For the Reissner–Nordström black hole in the PFDM background the topological number is $W=0$. A generate point is found  at \cite{rtop2PFDM}
\begin{equation}
\beta_c = 4\pi \frac{\left(-\alpha+\sqrt{3Q^2+\alpha^2}\right)^3}{2a^2-\alpha\left(-\alpha +\sqrt{3Q^2+\alpha^2}\right)}.
\end{equation}
For $\beta < \beta_c$, no black hole state is present. However, for the other values, $\beta_c < \beta$, there are two black hole states: small black holes with winding number $1$, and large black holes with winding number $-1$, representing thermodynamically stable and unstable states, respectively. However, the topological number is always zero, regardless of the values of $\beta$, the black hole parameter $\alpha$, or $Q$.

For the Reissner–Nordström black hole in the PFDM background case, the contours $\Phi_i$ that map the change in the components $(\phi^{r_h}, \phi^\Theta)$ of vector field $\phi$ as the contours $C_i$ in Fig. \ref{VFchKiselevBH} are traversed are plotted in Fig. \ref{F2ChKiselevBH}. The zero point of the field $\phi$ is at the origin, and contour with a positive winding number in Fig. \ref{VFchKiselevBH} is mapped to counterclockwise loop, whereas those with a negative winding number is mapped to clockwise loop in Fig. \ref{VFchKiselevBH}.
So, this confirms that for the Reissner–Nordström black hole in the PFDM background, the winding numbers of the first and second zero points are $+1$ and $-1$, respectively. The positive winding number $+1$ corresponds to the black hole state with positive heat capacity, and the negative winding number $-1$ corresponds to a black hole state with negative heat capacity. If additional black hole states are present, they must emerge in pairs, ensuring the topological number remains zero. 

The universal thermodynamical behavior is summarized as follows: in the low-temperature limit, $\beta \to \infty$, the system features an unstable large black hole and a stable small black hole, whereas in the high-temperature limit, $\beta \to 0$, no black hole state is present. Furthermore, the Reissner–Nordström black hole in the PFDM background belongs to a specific topological class $W^{0+}$. 
\begin{figure}
\centering
\includegraphics[width=7.5cm,height=7.5cm]{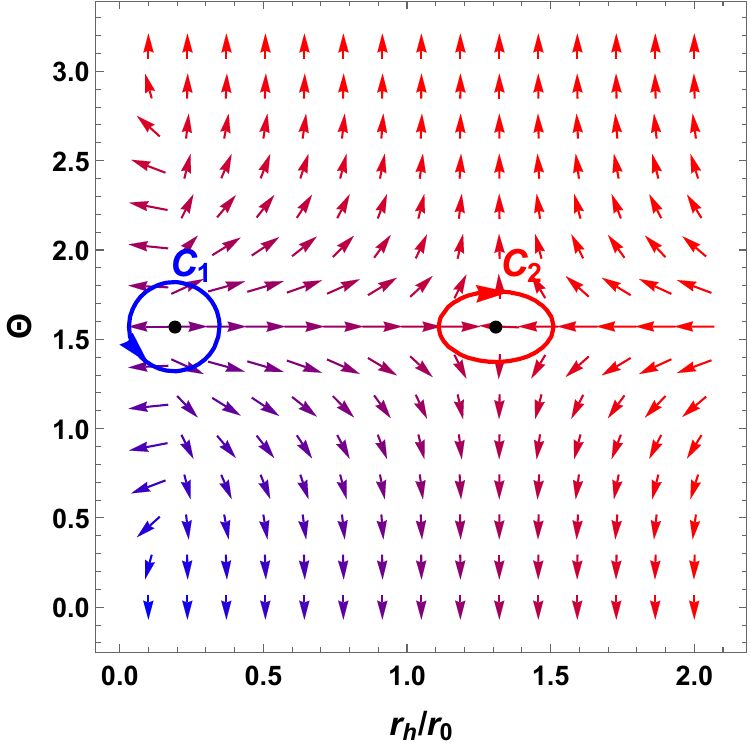}
\caption{The $n$-vector field on a portion of the $r_h$-$\Theta$ plane plotted for the Kerr-Newman black hole in PFDM background with $\alpha=1/2$, $a=1/4$, $Q=1/4$ and $\tau=4\pi$. The zero points of $\phi$ are marked with black dots and are located at $(0.19,\pi/2)$ and $(1.31,\pi/2)$.}
\label{VFKNKiselevBH}
\end{figure}
\begin{figure}
\centering
\includegraphics[width=7.5cm,height=7.5cm]{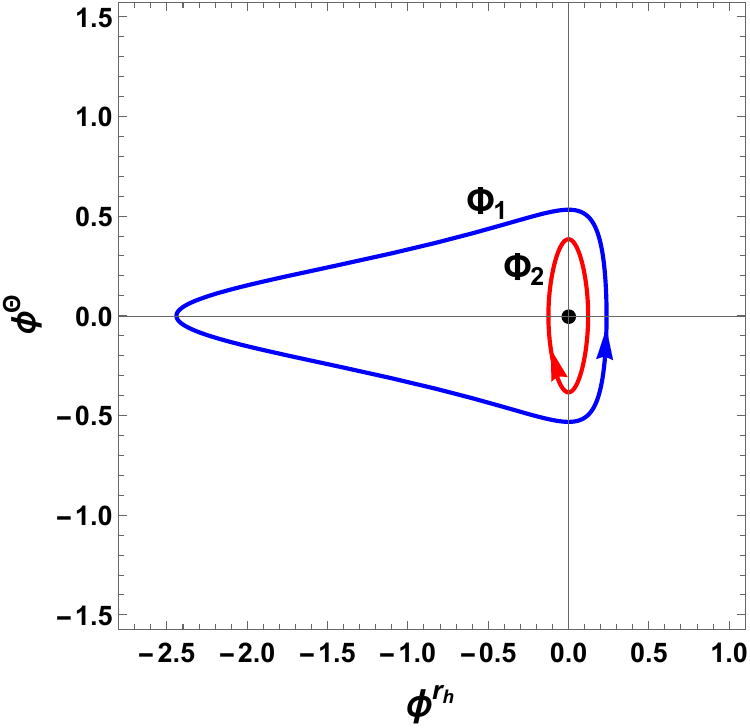}
\caption{Contours $\Phi_1$ and $\Phi_2$ represent the changes in the components of the vector field $\phi$ as the contours $C_1$ and $C_2$ in Fig. \ref{VFKNKiselevBH} are traversed for a Reissner–Nordström black hole in PFDM. The black dots indicate the zero points of the vector field, and the arrows show the direction of rotation of the vector $\phi$ while traversing the contours $C_1$ and $C_2$. The changes in the components of the field along the closed loop cause $\Phi_1$ to rotate in a counterclockwise direction, corresponding to a winding number of $1$, whereas along loop $\Phi_2$, it rotates in a clockwise direction, corresponding to a winding number of $-1$.}
\label{F3KNPFDMBHWN}
\end{figure}

\section{Kerr-Newman Black Hole in Dark Matter Background}

In this section, we discuss the effect of PFDM on topological classes of Kerr-Newman spacetime. The line element of the Kerr-Newman black hole in the dark matter background is given as \cite{KerrDM}
\begin{eqnarray}\label{LEK}\notag
ds^{2} &=&-\frac{\Delta_r}{\Sigma}\left(dt-a\sin^2\theta d\phi\right)^2+\frac{\Sigma }{\Delta_r }dr^{2}+\Sigma d\theta ^{2}\\
&+&\frac{ \sin^2\theta}{\Sigma} \left[adt-\left(r^2+a^2\right)d\phi\right]^2 ,
\end{eqnarray}%
where%
\begin{eqnarray}
\Delta_r &=& r^{2}-2mr+a^{2}+Q^2+\alpha r\ln \left( \frac{r}{|\alpha |}\right), \\
\Sigma&=& r^{2}+a^{2}\cos ^{2}\theta .
\end{eqnarray}%
Here, the thermodynamic quantities to investigate the topological classification of the Kerr-Newman black hole in PFDM can be presented as \cite{MJP}
\begin{eqnarray}
M&=&m, \quad S=\pi\left(r^2_h+a^2\right),\\
T&=&\frac{r_h}{4\pi\left(r^2_h+a^2\right)}\left(1-\frac{a^2+Q^2}{r^2_h}+\frac{\alpha}{r_h}\right).
\end{eqnarray}
This black hole minimal size Kerr-Newman black hole in PFDM background same as that of the the Reissner–Nordström
black hole in the PFDM background given by Eq. (\ref{extremal_radius}) which represents the horizon of an extremal black hole which is formed when the rotation parameter $a$, charge $Q$ and the mass parameter $m$ of the black hole satisfy the relation
\begin{eqnarray}
    a^2+Q^2=k^2_c.
\end{eqnarray}
The asymptotic behavior of the defect curve $\beta(r_h)$ at the boundaries is as follows:
\begin{eqnarray}
    \beta(r_m)=\infty \quad \text{and} \quad \beta(\infty)=\infty.
\end{eqnarray}
The generalize off-shell free energy for Kerr-Newman black hole in PFDM background is
\begin{eqnarray}\label{FRN}
\mathcal{F}&=&\frac{1}{2}\left[r_h+\frac{a^2+Q^2}{r_h}+\alpha \ln\left( \frac{r_h}{|\alpha |}\right)\right]-\frac{\pi (r_h^2+a^2)}{\tau},
\end{eqnarray}
and the components of the vector field $\phi$ are evaluated as
\begin{eqnarray}\label{phiKN}
\phi^{r_h} &=& \frac{1}{2}\left(1+\frac{\alpha}{r_h}-\frac{a^2+Q^2}{r^2_h}\right) - \frac{2\pi r_h}{\tau}, \\
\phi^\Theta &=& -\cot{\Theta} \csc{\Theta}.
\end{eqnarray}
The components of the vector field $\phi$ given by Eq. \eqref{phiKN} for the Kerr-Newman black hole in the PFDM background have the same form as those obtained for the Reissner–Nordström black hole in the PFDM background, as given by Eq. \eqref{phiRN}, provided $Q^2 \to a^2 + Q^2$. The graphical study of the Kerr-Newman black hole in the PFDM background, presented in Figs. \ref{VFKNKiselevBH} and \ref{F3KNPFDMBHWN}, is also the same. Therefore, we can conclude that the Kerr-Newman (and similarly, Kerr) black holes in the PFDM background belong to the same universal thermodynamical class, $W^{0+}$, as the Reissner–Nordström black hole in PFDM, and the thermodynamical states of these black holes are summarized in Table \ref{table1} and \ref{table3}.
\begin{table}[h]
    \centering
    \begin{tabular}{|l|c|c|c|c|c|}
        \hline
        \textbf{Black hole solutions }& ${I_1}$ & $I_2$ & ${I_3}$ & ${I_4}$ & ${W}$ \\
        \hline
        Schwarzschild black hole in PFDM & $\leftarrow$ & $\uparrow$ & $\rightarrow$ & $\downarrow$ & -1 \\
        Reissner–Nordström black hole in PFDM& $\leftarrow$ & $\uparrow$ & $\leftarrow$ & $\downarrow$ & 0 \\
        Kerr black hole in PFDM& $\leftarrow$ & $\uparrow$ & $\leftarrow$ & $\downarrow$ & 0 \\
        Kerr-Newman black hole in PFDM& $\leftarrow$ & $\uparrow$ & $\leftarrow$ & $\downarrow$ & 0 \\
        \hline
    \end{tabular}
    \caption{Comparison of black hole solutions and topological numbers.}
    \label{table1}
\end{table}

\section{Schwarzschild-AdS Black Hole in Dark Matter Background}

\begin{figure}
\centering
\includegraphics[width=7.5cm,height=7.5cm]{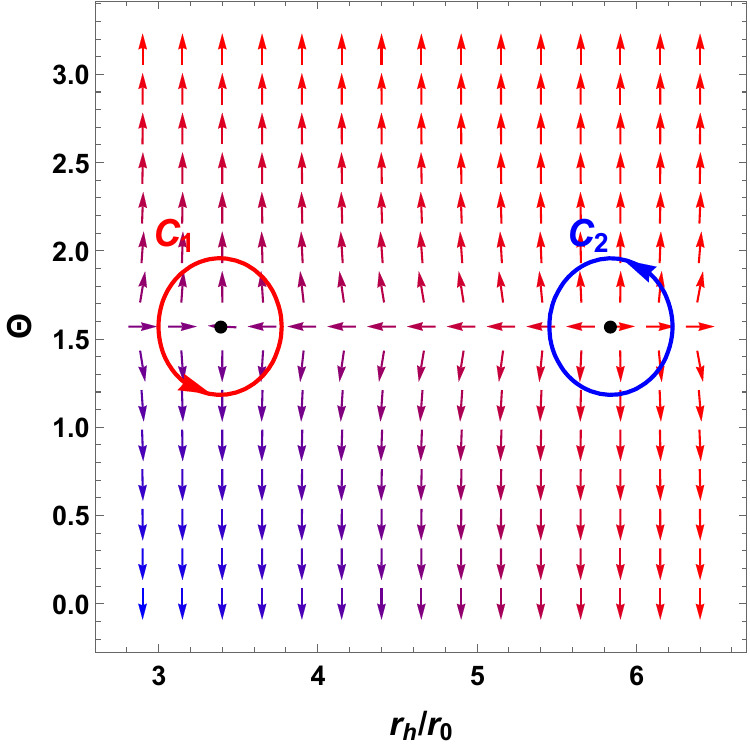}
\caption{The $n$-vector field on a portion of the $r_h$-$\Theta$ plane plotted for the Schwarzschild black hole in PFDM background with $\alpha=1/5$,  $P=0.0022$ and $\tau=8\pi$. The zero points of $\phi$ are marked with black dots and are located at $(3.39,\pi/2)$, and $(5.83,\pi/2)$.}
\label{VFSAdSBHinPFDM}
\end{figure}

Now we discuss the topological classes of the Schwarzschild-AdS black hole in PFDM background whose line element can be written as \cite{KerrDM} 
\begin{eqnarray}\label{LESAdSPFDM}
ds^2 &=&-f(r)dt^2+\frac{1}{f(r)}dr^2+r^2\left(d\theta^2+\sin^2\theta d\phi^2\right),
\end{eqnarray}
where
\begin{eqnarray}
 f(r)=1-\frac{2m}{r}-\frac{\Lambda}{3}r^2+\frac{\alpha}{r}\ln\left(\frac{r}{|\alpha|}\right).
\end{eqnarray}
Here $\Lambda<0$ denotes the cosmological constant. For all values of the black hole parameters, the Schwarzschild-AdS black hole in PFDM background has a single horizon, represented by $r_h$, and in this case, the minimal radius $r_m$ of the black hole is zero. The mass, entropy and the Hawking temperature of the Schwarzschild-AdS black hole in PFDM background is given by \cite{MJP}
\begin{eqnarray}
M&=&m, \quad  S=\pi r_h^2,\\
T&=&\frac{1}{4\pi r_h}\left[1-\Lambda r^2_h+\frac{\alpha}{r_h}\right].
\end{eqnarray}
The asymptotic behavior of the defect curve $\beta(r_h)$ at the boundaries is as follows:
\begin{eqnarray}
    \beta(r_m)=0 \quad \text{and} \quad \beta(\infty)=0.
\end{eqnarray}
These thermodynamic quantities lead to the off-shell free energy, and the corresponding components of the $n$-vector field $\phi$ for the Schwarzschild-AdS black hole have the form:
\begin{equation}
\mathcal{F}=\frac{1}{6} \left[3r_h+8 \pi  P r_h^3+3 \alpha  \ln \left(\frac{r_h}{|\alpha| }\right)\right]-\frac{\pi  r_h^2}{\tau },
\end{equation}
and 
\begin{eqnarray}
   \phi^{r_h}&=&\frac{1}{2}\left(1+\frac{\alpha}{r_h}  +8\pi P r_h^2\right)  -\frac{2 \pi  r_h}{\tau},\\
   \phi^\Theta&=&-\cot\Theta \csc\Theta.
\end{eqnarray}
where $P=-\Lambda/8\pi$ representing the pressure parameter. 
The vector field $\phi$ for the Schwarzschild-AdS black hole in the PFDM background, is shown in Fig. \ref{VFSAdSBHinPFDM}. The figure shows that, unlike the Schwarzschild black hole in PFDM, the Schwarzschild-AdS black hole in PFDM has two zero points, represented by black dots. The asymptotic behavior of the vector field $\phi$ at the boundaries $I_i$ is shown in Table \ref{table2}. It is observed that as $r_h \to r_m$ or if $r \to \infty$, the vector field plot of $\phi$ indicates a rightward direction, suggesting its universal thermodynamical class.

\begin{figure}
\centering
\includegraphics[width=7.5cm,height=7.5cm]{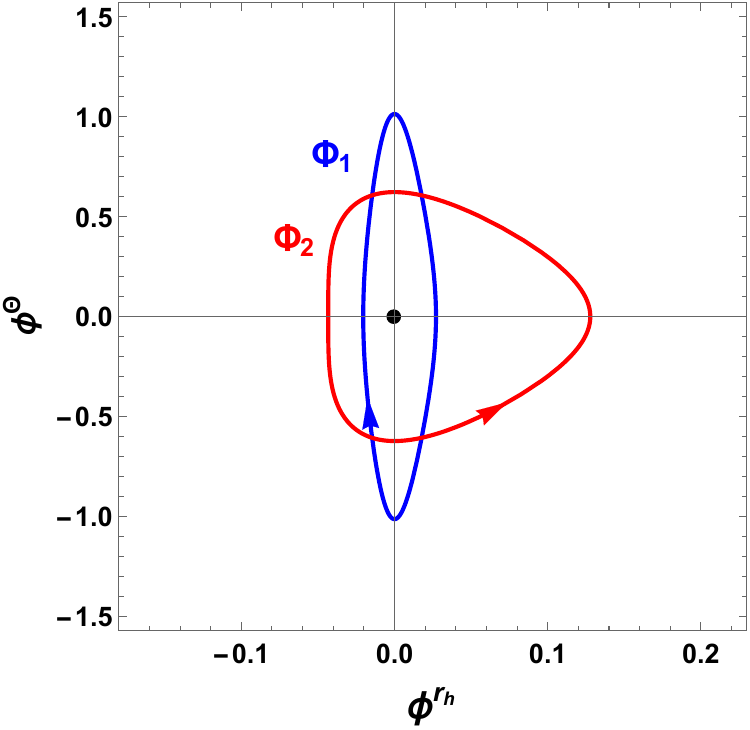}
\caption{Contours $\Phi_i$ represent the changes in the components of the vector field $\phi$ as the contours $C_i$ in Fig. \ref{VFSAdSBHinPFDM} are traversed for a Schwarzschild-AdS black hole in PFDM. The black dots indicate the zero points of the vector field, and the arrows show the direction of rotation of the vector $\phi$ while traversing the contours $C_i$. The changes in the components of the field along the closed loop cause $\Phi_1$ and to rotate in a clockwise direction, corresponding to a winding number of $-1$, whereas along loop $\Phi_2$, it rotates in a counterclockwise direction, corresponding to a winding number of $1$.}
\label{F2SAdSPFDMBH}
\end{figure}

For the Schwarzschild-AdS black hole in the PFDM background, the contours $\Phi_i$ that map the change in the components $(\phi^{r_h},\phi^{\Theta})$ of the vector field $\phi$ as the contours $C_i$ in Fig. \ref{VFSAdSBHinPFDM} are traversed are plotted in Fig. \ref{F2SAdSPFDMBH}. The zero point of the field $\phi$ is at the origin, and contours with negative winding number are mapped to clockwise loop, whereas with a positive winding number is mapped to counterclockwise loop. An annihilation point exists at $\beta=\beta_c$ (which depends on the parametric values of the black hole parameters). For $\beta < \beta_c$, there are two black hole states: small black holes with winding number $-1$, and large black holes with winding number $1$, representing thermodynamically unstable and stable black holes, respectively. However, for $\beta_c < \beta$, no black hole state is present. The topological classes remain unaffected regardless of the values of $\beta$ or other black hole parameters.

The universal thermodynamic behavior is summarized as follows: In the low-temperature limit $\beta \to \infty$, there are no black hole states. In the high-temperature limit $\beta \to 0$, unstable small-sized and stable large-sized black holes exist. Furthermore, the Schwarzschild-AdS black hole in the PFDM background belongs to a specific topological class, $W^{0-}$.

\section{Kerr-AdS Black Hole in Dark Matter Background}
In this section, we will discuss the universal topological class of Kerr-AdS in PFDM. The line element of the Kerr-AdS black hole in the dark matter background is given as \cite{SBHDM,KerrDM}
\begin{eqnarray}\label{LE}\notag
ds^{2} &=&-\frac{\Delta_r}{\Xi\Sigma}\left(dt-a\sin^2\theta d\phi\right)^2+\frac{\Sigma }{\Delta }dr^{2}+\frac{\Sigma}{\Delta_\theta} d\theta ^{2}\\
&+&\frac{\Delta_\theta \sin^2\theta}{\Xi\Sigma} \left[adt-\left(r^2+a^2\right)d\phi\right]^2 ,
\end{eqnarray}%
where%
\begin{eqnarray}\notag
\Delta_r &=& r^{2}-2mr+a^{2}-\frac{\Lambda}{3}r^2\left(r^2+a^2\right)+\alpha r\ln \left( \frac{r}{|\alpha |}\right), \\\notag
\Delta_\theta &=&1+\frac{\Lambda}{3}a^2\cos^2\theta, \\
\Sigma&=& r^{2}+a^{2}\cos ^{2}\theta, \quad \Xi=1+\frac{\Lambda}{3}a^2.
\end{eqnarray}%
Here $m$ and $a$ denote the mass and angular momentum per unit mass of the black hole while $\alpha$ and $\Lambda$ are the PFDM parameter and the cosmological constant, respectively. 
\begin{figure}
\centering
\includegraphics[width=7.5cm,height=7.5cm]{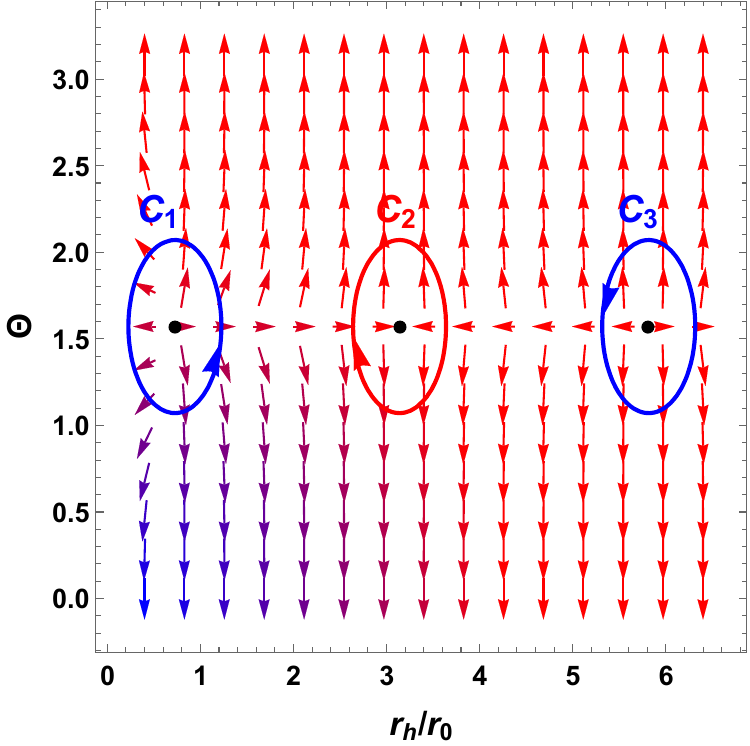}
\caption{The $n$-vector field on a portion of the $r_h$-$\Theta$ plane plotted for the \textcolor{black}{Kerr-AdS} black hole in PFDM background with $\alpha=1/5$, $a=1/2$, $Q=1/2$, $P=0.0022$ and $\tau=8\pi$. The zero points of $\phi$ are marked with black dots and are located at $(0.72,\pi/2)$, $(3.14,\pi/2)$ and $(5.82,\pi/2)$.}
\label{VFKNAdSPFDMBHWN}
\end{figure}

\begin{figure}
\centering
\includegraphics[width=7.5cm,height=7.5cm]{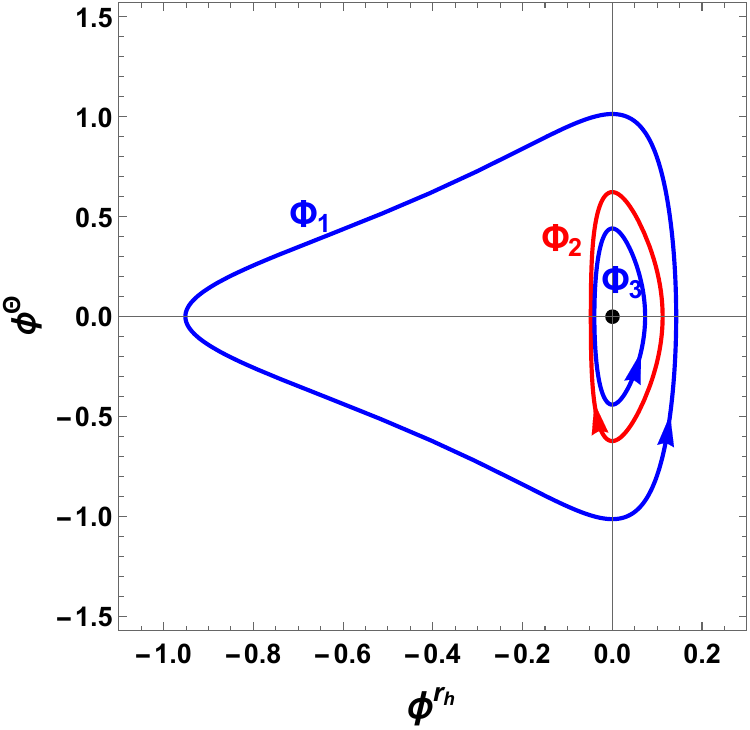}
\caption{Contours $\Phi_i$ represent the changes in the components of the vector field $\phi$ as the contours $C_i$ in Fig. \ref{VFKNAdSPFDMBHWN} are traversed for a Kerr-AdS black hole in PFDM. The black dots indicate the zero points of the vector field, and the arrows show the direction of rotation of the vector $\phi$ while traversing the contours $C_i$. The changes in the components of the field along the closed loop cause $\Phi_1$ and $\Phi_3$ to rotate in a counterclockwise direction, corresponding to a winding number of $1$, whereas along loop $\Phi_2$, it rotates in a clockwise direction, corresponding to a winding number of $-1$.}
\label{F2KNAdSPFDMBHWN}
\end{figure}

\begin{table}[h]
    \centering
    \begin{tabular}{|l|c|c|c|c|c|}
        \hline
        \textbf{Black hole solutions }& ${I_1}$ & $I_2$ & ${I_3}$ & ${I_4}$ & ${W}$ \\
        \hline
        Schwarzschild-AdS in PFDM & $\rightarrow$ & $\uparrow$ & $\rightarrow$ & $\downarrow$ & 0 \\
        Kerr-Ad BH in PFDM& $\rightarrow$ & $\uparrow$ & $\leftarrow$ & $\downarrow$ & 1 \\
        \hline
    \end{tabular}
    \caption{Comparison of black hole solutions and the topological numbers.}
    \label{table2}
\end{table}

\begin{table*}[ht]
    \centering
    \begin{tabular}{|c|l|c|c|c|c|c|}
        \hline
       \textbf{W Classes} & \textbf{BH solutions} & \textbf{Innermost} & \textbf{Outermost} & \textbf{Low $T$} & \textbf{High $T$} & \textbf{DP}   \\
        \hline
       $W^{1-}$ & Schwarzschild BH in PFDM & unstable & unstable & unstable large  & unstable large & in pairs \\
        \hline
        $W^{0+}$ & Reissner–Nordström BH in PFDM & stable & unstable & stable small + unstable large & no & one more GP  \\
        \hline
        $W^{0+}$ & Kerr BH in PFDM & stable & unstable & stable small + unstable large & no & one more GP  \\
        \hline
        $W^{0+}$ & Kerr-Newman BH in PFDM & stable & unstable & stable small + unstable large & no & one more GP  \\
        \hline
        $W^{0-}$ & Schwarzschild-AdS BH in PFDM & unstable & stable & no & unstable small + stable large  & one more AP  \\
        \hline
        $W^{1+}$ & Kerr-AdS BH in PFDM & stable & stable &  \begin{tabular}[c]{@{}c@{}} stable small\\+ unstable intermediate \end{tabular}  & \begin{tabular}[c]{@{}c@{}} stable large\\+ unstable intermediate \end{tabular} & in pair  \\
        \hline
    \end{tabular}
    \caption{Universal thermodynamic topological classifications of family of the black holes in PFDM background and their thermodynamical properties. DP, AP and GP are used to represent degenerate point, annihilation point, and generation
 point, respectively.}
    \label{table3}
\end{table*}
In the absence of PFDM ($\alpha =0$), the line element %
\eqref{LE} represents a Kerr-AdS black hole. The location of the black hole horizons can be obtained by solving the horizon equation 
\begin{equation}  \label{Horizneq}
\Delta_r =r^{2}-2mr+a^{2}-\frac{\Lambda}{3}r^2\left(r^2+a^2\right)+\alpha r\ln \left( \frac{r}{|\alpha |}\right) =0.
\end{equation}
The asymptotic behavior of the defect curve $\beta(r_h)$ at the boundaries is given by:
\begin{eqnarray}
    \beta(r_m) = \infty \quad \text{and} \quad \beta(\infty) = 0.
\end{eqnarray}
The mass and other thermodynamical quantities of Kerr-AdS black hole PFDM background are given by \cite{MJP}
\begin{eqnarray}
M&=&\frac{m}{\Xi^2}, \quad  S=\frac{\pi\left(r_h^2+a^2\right)}{\Xi},\\
T&=&\frac{r_h}{4\pi\Xi\left(r_h^2+a^2\right)}\left[1+\frac{\alpha}{r_h}-\frac{a^2}{r^2_h}-\frac{\Lambda}{3}\left(3r^2_h+a^2\right)\right].
\end{eqnarray}
Using these thermodynamical quantities, the generalized off-shell free energy of the system can be obtained as 
\begin{equation}
\mathcal{F}=\frac{1}{2 r_h\tau\left(3-8\pi a^2P\right)^2}
\Big[3 \left(r_h^2+a^2\right)\mathcal{G}+9 \alpha r_h \tau \ln \left(\frac{r_h}{\left| \alpha \right| }\right)\Big],
\end{equation}
where 
\begin{eqnarray}
    \mathcal{G}=2 \pi r_h \left(8 \pi P a^2 +4\tau P r_h-3\right)+ 3 \tau.
\end{eqnarray}
which gives the the components of $n$-vector field as
\begin{eqnarray}
   \phi^{r_h}&=&\frac{6\pi r^2_h\left\{8\pi a^2r_h P-3r_h+2\tau P\left(a^2+3r^2_h\right)\right\}}{2r^2_h\tau\left(3-8\pi a^2 P\right)^2}\nonumber\\
   &&+\frac{9\tau\left(r^2_h+\alpha r_h-a^2\right)}{2\tau r^2_h\left(3-8\pi a^2P\right)},\\
   \phi^\Theta&=&-\cot\Theta \csc\Theta.
\end{eqnarray}

The vector field $\phi$ for the Kerr-AdS black hole in the PFDM background is plotted in Fig. \ref{VFKNAdSPFDMBHWN}, which shows that there are three zero points of the vector field, denoted by black dots. These zero points divide the black hole configuration into a small, intermediate, and large black hole type, with one generation point and one annihilation point. Furthermore, the asymptotic behavior of the vector field at the boundaries $l_i$ is shown in Table \ref{table2}. The direction of the vector field at the boundary $I_1$ is rightward, while at the boundary $I_3$, it is leftward, suggesting its universal class.

For the Kerr-AdS black hole in PFDM, we have two critical values, $\tau_a$ and $\tau_b$. For $\tau < \tau_a$ or $\tau_b < \tau$, there is only black hole state with winding number $1$ representing large and small size black hole respectively . However, for $\tau_a < \tau < \tau_b$, three black hole states exist: small, intermediate, and large black holes. The contours $\Phi_i$, which map the changes in the components $\left(\phi^{r_h}, \phi^\theta\right)$ of the vector field $\phi$ as the contours $C_i$ shown in Fig. \ref{VFKNAdSPFDMBHWN} are traversed, are plotted in Fig. \ref{F2KNAdSPFDMBHWN}. These contours yield the winding numbers of the zero points as $1$, $-1$, and $1$, respectively. Thus, the small and large black holes have positive heat capacity and hence are stable, whereas the intermediate black holes have negative heat capacity and are consequently unstable.

The universal thermodynamic behavior is as follows: In the low-temperature limit $\beta \to \infty$, the system features stable small-sized and unstable intermediate-sized black holes. In the high-temperature limit $\beta \to 0$, there are unstable intermediate-sized and stable large-sized black holes. Thus, the Kerr-AdS black hole in the PFDM background belongs to the topological class $W^{1+}$.

\section{Conclusions}

In a study, Wei et al. \cite{rtop3unipaper}  proposed that all black holes can be thermodynamically classified into four universal topological classes: $W^{1-}$, $W^{0+}$, $W^{0-}$ and $W^{1+}$. These classification provides valuable insights into the thermodynamic behavior of small and large black holes in the low- and high-temperature limits. We have studied the Schwarzschild, Reissner-Nordström, Kerr-Newman, Schwarzschild-AdS and Kerr-AdS black holes in the PFDM background. The topological classification and thermodynamical stability of black holes in PFDM at low and high temperatures are summarized in Table. \ref{table3}.

The Schwarzschild black hole in the PFDM has one state with a negative winding number and, hence, a negative heat capacity. Thus, it belongs to the universal thermodynamical class $W^{1-}$, with thermodynamically unstable small and large black holes at both low and high temperature limits.  For the Reissner-Nordström, Kerr, and Kerr-Newman black holes in PFDM background, when $\beta<\beta_c$ there is no black hole state whereas for  $\beta_c < \beta$, there are two black hole states: a small black hole and a large black hole. In the low temperature limit, the small black holes have a winding number of $1$, and the large black holes have a winding number of $-1$, representing thermodynamically stable and unstable states, respectively. Furthermore, at high temperatures, no black hole state exists. The Reissner-Nordström, Kerr, and Kerr-Newman black holes in PFDM background all belong to the same thermodynamical class $W^{0+}$. 

Black holes in the AdS background exhibit different topological numbers compared to their counterparts that are not in the AdS background. The Schwarzschild-AdS black hole in PFDM background, for low-temperature limit, there is no black hole state whereas for high-temperature limit,
there are two black hole states with stable-small size and unstable-large size black holes. The Schwarzschild-AdS black hole in PFDM background belong to the universal thermodynamical class $W^{0-}$.

For the Kerr-AdS black hole in the PFDM background, for any chosen values of the black hole parameters, when $\beta < \beta_a$ or $\beta_b < \beta$, there exists only one black hole state, representing either a stable large or a stable small black hole. For $\beta_a < \beta < \beta_b$, the Kerr-AdS black hole in the PFDM background exhibits both generation as well as annihilation points, resulting in three black hole states: small, intermediate, and large. At low temperatures, there is a stable-small sized black hole and an unstable-intermediate-sized black hole, whereas in the high-temperature limit, there is a stable-large sized black hole and an unstable-small sized black hole. Furthermore, the Kerr-AdS black hole in the PFDM background belongs to the $W^{1+}$ universal thermodynamic class.

This study can be summarized as follows: the universal topological classes and the stability analysis based on these classes of black holes with PFDM background are consistent with those of black holes without PFDM background, indicating that PFDM does not alter the results \cite{rtop3unipaper,rtop4unitopo}.

\end{document}